\documentclass[sigconf]{acmart}
\setcopyright{none}
\acmISBN{}
\acmDOI{}
\acmConference[]{}{}{}
\makeatletter
\renewcommand{\footnotetextcopyrightpermission}[1]{}
\makeatother

\AtBeginDocument{%
  }

\usepackage{subcaption}
\usepackage{booktabs,tabularx}
\begin{document}

\title{Estimate Level Adjustment For Inference With Proxies Under Random Distribution Shifts}

\author{Steven Wilkins-Reeves}
\email{stevewr@meta.com}
\affiliation{%
  \institution{Central Applied Science, Meta}
  \city{New York}
  \state{NY}
  \country{USA}
}

\author{Alexandra N. M. Darmon}
\email{alexdarmon@meta.com}
\affiliation{%
  \institution{Central Applied Science, Meta}
  \city{Menlo Park}
  \state{CA}
  \country{USA}
}

\author{Deeksha Sinha}
\email{deekshasinha@meta.com}
\affiliation{%
  \institution{Central Applied Science, Meta}
  \city{Menlo Park}
  \state{CA}
  \country{USA}
}

\renewcommand{\shortauthors}{Wilkins-Reeves et al.}

\begin{abstract}
In many scientific domains, including experimentation, researchers rely on measurements of proxy outcomes to achieve faster and more frequent reads, especially when the primary outcome of interest is challenging to measure directly. While proxies offer a more readily accessible observation for inference, the ultimate goal is to draw statistical inferences about the primary outcome parameter and proxy data are typically imperfect in some ways. 
To correct for these imperfections, current statistical inference methods often depend on strict identifying assumptions (such as surrogacy, covariate/label shift, or missingness assumptions). These assumptions can be difficult to validate and may be violated by various additional sources of distribution shift, potentially leading to biased parameter estimates and miscalibrated uncertainty quantification.
We introduce an estimate-level framework, inspired by domain adaptation techniques, to empirically calibrate proxy-based inference. This framework models the proxy–primary metric discrepancy as a random effect at the parameter level, estimating its distribution from aggregated historical observations across past domains (e.g., experiments, time periods, or distinct segments). This method avoids the requirement for retaining individual-level response data. Additionally, this adjustment can be layered on top of existing proxy-correction methods (such as prediction-powered inference or importance weighting) to account for additional biases not addressed by those corrections. 
To manage uncertainty when the number of historical domains is limited, we provide both a method-of-moments estimator and a domain bootstrap procedure. Through extensive simulations, our adjusted intervals demonstrate improved calibration and more reliable coverage across various proxy procedures and deviations from standard covariate shift mechanisms. We further validate this approach using publicly available datasets and real-world experiments. Code for replicating simulations and experiments on public datasets are available at \url{github.com/facebookresearch/estimate-level-adjustment}. 
\end{abstract}

\received{1 Feb 2026}

\maketitle

\fancyhead[LE]{}
\fancyhead[RO]{}

\section{Introduction}
In many scientific and industrial settings, the \emph{primary} outcome of interest is difficult to measure. As a result, practitioners frequently substitute a more readily available \emph{proxy} outcome, such as a short-term behavioral metric, a surrogate endpoint, or a model prediction, and then make decisions using inference computed on that proxy. The difficulty is that statistical guarantees for the proxy (e.g., unbiasedness, variance estimates, confidence intervals) do not automatically translate into reliable inference for the primary metric. Even when a proxy is strongly correlated with the target, proxy-based confidence intervals can be systematically miscalibrated for the primary estimand due to residual bias and distribution shift.

This problem has been studied for decades under different names and assumptions. In medicine, surrogate endpoints are typically justified via strong causal requirements, often phrased as the surrogate \emph{mediating} the effect of treatment on the final outcome—and validated using extensive clinical trial evidence \cite{prentice1989surrogate,buyse2000validation}. In social sciences, where a single mediating surrogate may be hard to defend, researchers often construct surrogate \emph{indices} by combining multiple proxies \cite{heckman2006effects,athey2019surrogate}. In online experimentation and large-scale decision systems, proxy metrics are pervasive because primary outcomes can be delayed, rare, or expensive to observe \cite{gupta2019top}. Modern platforms also increasingly rely on model-based measurements—e.g., machine-assisted estimation of policy-violating content prevalence \cite{nguyen2020clara}.

Despite these advances, a persistent practical issue remains: \emph{even after} applying a correction method (or even when using a proxy that appears well-motivated), the final estimate may be biased, and this residual bias can induce systematic under-coverage of reported confidence intervals for the primary estimand. Moreover, individual-level historical data needed to re-fit or re-calibrate models may not be retained; in contrast, aggregate metric estimates and uncertainty summaries are often preserved.

We propose a generic adjustment procedure for confidence intervals computed from a proxy metric. Our setup allows the proxy estimator to be biased for the primary estimand and assumes that the proxy estimator's sampling error is asymptotically normal, an assumption aligned with standard large-sample approximations for aggregate metrics. We develop an approach to model and estimate the distribution of residual proxy bias without requiring individual-level historical data. Instead, our method uses only historical aggregate proxy estimates (and associated uncertainty summaries such as covariances), enabling scalable deployment in settings where only aggregate data are retained. Our adjustment procedure is applied as a post-processing step, which allows for its use on top of proxy estimates. 

At a high level, our method treats residual proxy bias as a latent variable and uses a method-of-moments estimator to infer its distribution from repeated historical measurements. This estimated bias distribution then induces an explicit inflation (or adjustment) of reported confidence intervals so that coverage is improved for the primary estimand. 

We begin by formalizing the proxy inference problem and provide an example of different proxy estimates and possible identifying assumptions. We then present our latent-bias modeling framework and method-of-moments estimator, and derive adjusted confidence intervals under this model, as well as a bootstrapped version to correct for latent model uncertainty. We illustrate the approach on simulated and real-data case studies, and conclude with limitations and directions for future work, highlighting the selective inflation of only biased confidence intervals and adaptability of the inflation to different choices of proxy metric.

\subsection{Related Work}
\label{subsec:related_work}
Our work relates to several strands of literature on inference and decision-making with proxies.
\paragraph{Surrogates and proxy outcomes.}
Surrogate endpoints in medicine have a long history and often rely on strong causal and validation assumptions \cite{prentice1989surrogate,buyse2000validation}. In the social sciences and policy evaluation, proxies and surrogate indices are frequently used to summarize multi-faceted outcomes when direct measurement is difficult \cite{heckman2006effects,athey2019surrogate}. In online experimentation and large-scale product decision-making, proxies are ubiquitous due to difficult to measure outcomes \cite{gupta2019top}, and proxy construction may involve operational constraints such as measurement noise, limited labels, or changing instrumentation \cite{nguyen2020clara,sarig2025mind}. While surrogates typically rely strongly on the assumptions used to identify the outcome, our method does not require such stringent identifying assumptions and instead adjusts estimates based on historical bias. 

\paragraph{Inference with predictions and prediction-powered inference.}
The statistical challenges of using predicted outcomes for inference were recognized in early work \cite{wang2020methods}. The prediction-powered inference (PPI) framework of \cite{angelopoulos2023prediction} and its extensions \cite{angelopoulos2023ppi++,bashari2025synthetic} provide general procedures with formal guarantees under assumptions on how labeled and unlabeled data relate (e.g., missing-completely-at-random and structured missingness) \cite{chen2025unified,ji2025predictions,kluger2025prediction}, drawing connections to semiparametric efficiency \cite{tsiatis2006semiparametric}. Though a simple and robust correction, these methods assume that the relationship between the proxy and the primary is stable across labelled and unlabelled domains. We will illustrate an example where such estimators fail and how our procedure can correct the estimates.

\paragraph{Population-level inference from model properties.}
In prevalence estimation and related tasks, population-level inference can sometimes be supported by known model properties such as sensitivity/specificity \cite{rogan1978estimating}. However, when labels are biased or sampling is ill-defined, estimating these properties for the target population requires assumptions akin to missing data conditions. Calibration-style guarantees—such as multicalibration and multiaccuracy \cite{hebert2018multicalibration,kim2019multiaccuracy}, and their ``universal adaptability'' implications \cite{kim2022universal}—can also support valid inference across populations under appropriate assumptions. Related calibration methods have been applied to prevalence estimation \cite{wu2024calibrate}, under assumptions of either covariate or label shift, and relate these to the causal relationships of the primary outcome and the related features \cite{scholkopf2012causal}. Complementary recent work develops methods for learning and choosing proxies for decision-making and experimentation \cite{bibaut2024learning,tran2024inferring,tripuraneni2024choosing}.

\paragraph{Empirical Bayes and shrinkage for uncertainty calibration.}
Methodologically, our approach shares most in common with empirical Bayes methodologies.  \cite{robbins1964empirical,efron2001empirical,ignatiadis2022confidence,ignatiadis2023Empirical} in that we estimate a latent distribution (here, residual proxy bias) and use it to improve uncertainty quantification. Empirical Bayes methods have been deployed in experimental contexts—e.g., pooling across experiments \cite{ejdemyr2024estimating} and mitigating the winner's curse \cite{lee2018winner,thaler1988anomalies,andrews2024inference}, however, these are not the focus of this paper. Our working model to correct the estimate of the latent bias between proxies is most similar in structure to the meta-analytic model of \cite{dersimonian1986meta}. To our knowledge, directly using an empirical-Bayes style latent distribution to adjust confidence intervals for proxy-based inference—especially when only historical \emph{aggregate} summaries are available—has not been previously studied.

\subsection{Our Contributions}
\label{subsec:contributions}
The contributions of this paper are:
\begin{enumerate}
    \item \textbf{A generic framework for the biases of proxy metrics.}
    We present a framework for characterizing the bias inherent in proxy estimators relative to the primary estimand. Assuming the joint normality of the sampling distributions of the estimators, a relatively mild condition, we show that the residual bias can be formulated as a latent variable problem.
    \item \textbf{A simple adjustment procedure for inflating confidence intervals.}
    We introduce a working model and a closed-form method-of-moments estimator for the bias distribution, which relies solely on historical \emph{aggregate} proxy estimates and their covariance structure. This approach is scalable, time and compute efficient and suitable for scenarios where individual-level data are unavailable due to some constraints. For cases with a limited number of historical domains, we provide a bootstrapped version to account for this limitation.
    \item \textbf{Practical guidance and empirical validation, including selective inflation.}
    Through extensive simulations and real-data case studies, we demonstrate that our method improves the coverage of adjusted intervals over time. Notably, our approach inflates intervals only when diagnostics indicate that the proxy would otherwise undercover. We also highlight situations where the bootstrap approach offers advantages over the standard variant.

\end{enumerate}

\section{Methods}

We formalize a multi-domain setting in which the \emph{primary} outcome is unavailable in a target domain, and practitioners must rely on a \emph{proxy} metric (potentially corrected using labeled source domains). Our goal is to construct confidence intervals for the target-domain primary metric using the proxy. We proceed by (i) defining the environment and target parameters, (ii) introducing an estimate-level latent-bias model that uses only aggregate estimates and covariances, and (iii) describing an empirical leave-one-domain-out (LOO) evaluation protocol for proxy-based interval procedures.
\subsection{Environment}
\label{subsec:environment}
For each domain $k \in \{1,\dots,K\}$, we consider i.i.d.\ unit-level data
\[
O_{i} = (Y_{i}, Y^{\ast}_{i}, X_{i}) \sim P_k \in \mathcal{P}, \qquad i=1,\dots,n_k,
\]
where $Y_{i}$ is the \emph{primary} outcome of interest, $Y^{\ast}_{i}$ is a \emph{proxy} outcome, and $X_{i}$ denotes auxiliary covariates used for estimation and/or adjustment. 
We let $N_k$ denote the set of indices denoting observations in domain $k$ and let $n_k = |N_k|$ denote this total. We use the term \textbf{domain} to denote a segment of the data-generating process under which the joint distribution of $(Y,Y^\ast,X)$ may change. For example, a computer vision model's predictive behavior may differ across lighting conditions, inducing domain-dependent proxy bias. 

We assume that domains $k=1,\dots,K-1$, which we refer to as the training domains, contain \emph{complete} data $(Y,Y^\ast,X)$, while the target domain $k=K$ contains only $(Y^\ast,X)$ (i.e., $Y$ is unobserved). Our goal is to infer a target parameter
\[
\theta_K := \Psi(P_K) \in \mathbb{R},
\]
for a functional $\Psi:\mathcal{P}\to\mathbb{R}$; we refer to $\theta_K$ as the \textbf{primary metric}. Let $P^\ast_K$ denote the marginal distribution of $(Y^\ast,X)$ in the target domain. A \textbf{proxy metric} is a  functional
\[
\theta^\ast_K := \Psi^\ast(P^\ast_K, P_{-K}) \in \mathbb{R},
\qquad
P_{-K} := (P_1,\dots,P_{K-1}),
\]
which may depend on target proxy data and (complete) source-domain distributions. This framework includes common targets such as population means, quantiles, and treatment effects.\footnote{For causal estimands, $O_{i}$ can be extended to include treatment assignments and potential outcomes, under standard identification assumptions (e.g., randomization or unconfoundedness).}
We refer to estimators of $\theta_K$ and $\theta^\ast_K$ as the \emph{primary} and \emph{proxy} estimators, respectively. To adjust the intervals, we will not need to observe directly the outcomes, $O_i$, but rather the estimates themselves. Though letting $\Psi^* = \Psi$ and using the proxy directly in place of the primary outcomes are included in this framework, we allow for more general versions of proxies, which leverage the training domains to augment the estimate. We provide an illustration of this below.  

\paragraph{Example: covariate-shift proxy correction.}
Consider a simplified setting with one labeled source domain ($k=1$) and one target domain ($k=2$). A simple modeling assumption is \emph{covariate shift}\footnote{Analogous proxy metrics can be constructed under label-shift assumptions \cite{chan2005word,lipton2018detecting}.} on the residual $Y-Y^\ast$ \citep{shimodaira2000improving,sugiyama2007direct}, namely
\begin{align}
P_1(X) \neq P_2(X),
\qquad
P_{1}(Y-Y^\ast \mid X=x) = P_{2}(Y-Y^\ast \mid X=x).
\label{eq:cov_shift_residual}
\end{align}
Define the density ratio weights
\[
w_{12}(x) := \frac{p_2(x)}{p_1(x)},
\]
where $p_k$ denotes the density (or probability mass function) of $X$ under $P_k$. When the covariate shift assumption holds, one can construct a proxy metric for the target mean $\theta_2 = \mathbb{E}[Y\mid K=2]$
\begin{equation*}
\theta_2^\ast
:= \mathbb{E}[Y^\ast \mid K=2]
\;+\;
\mathbb{E}\!\left[(Y-Y^\ast) w_{12}(X)\,\middle|\,K=1\right]
\;=\;
\Psi^\ast(P_2^\ast,P_1). 
\end{equation*}
If equation~\eqref{eq:cov_shift_residual} holds and $w_{12}$ is correctly specified, then $\theta_2^\ast=\theta_2$. Importantly, $\theta_2^\ast$ is well-defined even when \eqref{eq:cov_shift_residual} does not hold; our focus is precisely on constructing uncertainty quantification that remains valid in the presence of \emph{residual} (domain-varying) bias.

\subsection{A Model for Proxy Bias}
\label{subsec:estimate_level}
We now introduce an estimate-level model for residual proxy bias across domains. For each labeled domain $k \in \{1,\dots,K-1\}$, let $(\widehat{\theta}_k,\widehat{\theta}_k^\ast)$ estimate $(\theta_k,\theta_k^\ast)$. We assume a joint asymptotic normal approximation of the form
\begin{equation}
\begin{pmatrix}
\widehat{\theta}_k \\
\widehat{\theta}_k^\ast
\end{pmatrix}
\;\sim\;
\mathcal{N}\!\left(
\begin{pmatrix}
\theta_k \\
\theta^*_k
\end{pmatrix},
\Sigma_k
\right),
\qquad
\Sigma_k :=
\begin{pmatrix}
\sigma_{0,k}^2 & \sigma_{0\ast,k} \\
\sigma_{0\ast,k} & \sigma_{\ast,k}^2
\end{pmatrix}.
\label{eq:joint_asymp_normal}
\end{equation}
This assumption is standard for many regular, asymptotically linear functionals (e.g., means and quantiles); see \cite{van2000asymptotic} for regularity (smoothness) conditions under which such approximations hold for $\Psi,\Psi^\ast$. We treat $\Sigma_k$ as known or consistently estimable (e.g., via influence-function based variance estimators or other large-sample formulas).

We model the discrepancy between proxy and primary \emph{parameters} across domains via a random-effects mechanism. In general, we posit a conditional distribution $G(\cdot \mid \theta_k)$ such that
\begin{equation}
\theta_k^\ast \,\big|\, \theta_k \sim G(\cdot \mid \theta_k).
\label{eq:G_general}
\end{equation}
A simple and practically useful working model is additive Gaussian bias:
\begin{equation}
\phi_k := \theta_k^\ast - \theta_k \sim \mathcal{N}(\rho,\gamma^2),
\qquad k=1,\dots,K,
\label{eq:normal_error}
\end{equation}
with unknown hyperparameters $(\rho,\gamma^2)$. This specification is structurally similar to classical random-effects meta-analysis \cite{dersimonian1986meta}, but our objective differs: we use the across-domain variability in $\phi_k$ to adjust (inflate) proxy-based confidence intervals so they reflect both within-domain sampling error and between-domain proxy bias variability.

\subsection{Evaluation of Proxy Interval Procedures}
\label{subsec:evaluation}
Our goal is to construct a confidence interval for the target-domain primary metric $\theta_K$ using (i) the target-domain proxy estimator $\widehat{\theta}_K^\ast$ (and its estimated variance) and (ii) historical information from domains $1,\dots,K-1$. We denote the resulting interval by $\mathcal{I}^\ast_{K,1-\alpha}$.
The classical (uniform) frequentist coverage (over all possible parameter values $\Theta$) requirement is
\begin{equation}
\inf_{\theta_K \in \Theta}\;
\mathbb{P}\!\left(
\theta_K \in \mathcal{I}_{K,1-\alpha}
\right)
\ge 1-\alpha.
\label{eq:frequentist_coverage}
\end{equation}
When residual proxy bias $\phi_K$ can vary arbitrarily across domains, guarantees of the form \eqref{eq:frequentist_coverage} can be unattainable without additional restrictions, since any one target-domain bias may differ from what was seen in the training set.
Instead, we target a weaker notion that averages over the latent bias mechanism in \eqref{eq:G_general}--\eqref{eq:normal_error}. One natural criterion is \textbf{average frequentist coverage}:
\begin{align}
\inf_{\theta_K \in \Theta}\;
\mathbb{P}\!\left(
\theta_K \in \mathcal{I}^\ast_{K,1-\alpha}(\widehat{\theta}_K^\ast,\mathcal{D}_{-K})
\;\middle|\; \theta_K
\right)
\ge 1-\alpha,
\label{eq:avg_freq_coverage}
\end{align}
where the probability integrates over the latent randomness in $\theta_K^\ast \mid \theta_K$ (and over the sampling distribution of $\widehat{\theta}_K^\ast$), and $\mathcal{D}_{-K}$ denotes the historical data from domains $1,\dots,K-1$ used by the procedure.
\paragraph{Leave-one-domain-out (LOO) calibration diagnostic.}
Because $\theta_k$ is unknown, coverage cannot be directly checked in a single historical domain. We therefore propose an empirical LOO diagnostic: for each $k \le K-1$, treat domain $k$ as if it were the target (i.e., hide $Y$), construct a proxy-based interval $\mathcal{I}^{\ast,-k}_{k,1-\alpha}$ using $(\widehat{\theta}_k^\ast,\widehat{\Sigma}_k)$ and the remaining domains $\{1,\dots,K-1\}\setminus\{k\}$, and compare it to a \emph{primary} interval $\mathcal{I}_{k,1-\alpha}$ built from $\widehat{\theta}_k$.
A conservative, observable proxy for non-miscalibration is whether the two intervals overlap:
\[
\mathbf{1}\!\left\{\mathcal{I}^{\ast,-k}_{k,1-\alpha} \cap \mathcal{I}_{k,1-\alpha} \neq \emptyset \right\}.
\]
Aggregating across historical domains yields the empirical overlap rate
\begin{equation}
\widehat{g}(\alpha)
:=
\frac{1}{K-1}\sum_{k=1}^{K-1}
\mathbf{1}\!\left\{\mathcal{I}^{\ast,-k}_{k,1-\alpha} \cap \mathcal{I}_{k,1-\alpha} \neq \emptyset \right\},
\label{eq:overlap_rate}
\end{equation}
which provides a necessary (though generally conservative) diagnostic: if $\theta_k \in \mathcal{I}_{k,1-\alpha}$ and $\mathcal{I}^{\ast,-k}_{k,1-\alpha}$ overlaps $\mathcal{I}_{k,1-\alpha}$, then $\mathcal{I}^{\ast,-k}_{k,1-\alpha}$ is consistent with plausible values of $\theta_k$ at level $1-\alpha$. We use this diagnostic in our experiments to compare proxy interval procedures and to motivate adjustment strategies that expand proxy intervals primarily when historical LOO evidence suggests under-coverage.

\section{Estimation and Inference for the Random-Bias Model}
\label{sec:randombias}
Throughout, we adopt the additive Gaussian residual-bias model~\eqref{eq:normal_error} and the joint asymptotic approximation~\eqref{eq:joint_asymp_normal}. For each labeled (historical) domain $k=1,\dots,K-1$, define the within-domain proxy--primary difference at the \emph{estimate} level,
\[
d_k := \widehat{\theta}_k^\ast - \widehat{\theta}_k.
\]
Under~\eqref{eq:joint_asymp_normal}, the sampling error in $d_k$ is approximately normal with variance
\[
\widetilde{\sigma}_k^2
:= \text{Var}(\widehat{\theta}_k^\ast-\widehat{\theta}_k)
= \sigma_{\ast,k}^2 + \sigma_{0,k}^2 - 2\sigma_{0\ast,k}.
\]
Combining this with $\Delta_k \sim \mathcal{N}(\rho,\gamma^2)$ yields the marginal approximation
\[
d_k \; \sim\; \mathcal{N}\!\left(\rho,\; \gamma^2 + \widetilde{\sigma}_k^2\right),
\qquad k=1,\dots,K-1.
\]
Let $\overline{d}:=\frac{1}{K-1}\sum_{k=1}^{K-1} d_k$. A simple method-of-moments estimator is
\begin{align}
\widehat{\rho}
&:= \overline{d}
= \frac{1}{K-1}\sum_{k=1}^{K-1}\left(\widehat{\theta}_k^\ast-\widehat{\theta}_k\right),
\label{eq:mom_rho}
\\
\widehat{\gamma}^2
&:= \frac{1}{K-1}\sum_{k=1}^{K-1}\left(d_k-\overline{d}\right)^2
\;-\;
\frac{1}{K-1}\sum_{k=1}^{K-1}\widetilde{\sigma}_k^2.
\label{eq:mom_gamma}
\end{align}
In implementation, we truncate at zero to avoid negative estimates due to finite-sample noise:
\[
\widehat{\gamma}^2 \leftarrow \max\{0,\widehat{\gamma}^2\}.
\]
Heuristically, $\widehat{\gamma}^2=0$ indicates that, at the resolution of the within-domain estimation error ($\widetilde{\sigma}^2_k$), there is insufficient evidence of \emph{between-domain} variability in residual proxy bias beyond what would be expected from sampling variability alone. As an extension, in the appendix in section~\ref{subsec:contextual} we illustrate how add additional contextual information in the adjustment, such as fine-tuning $(\rho, \gamma^2)$ parameters across time. 

\subsection{Inference in the target domain}
\label{subsec:inference}
We now construct confidence intervals for $\theta_K$ in the target domain, using the target proxy estimate $\widehat{\theta}_K^\ast$ and historical domains to characterize residual bias.
\paragraph{Large $(K-1)$: plug-in Wald interval.}
When the number of historical domains is large, we treat $(\rho,\gamma^2)$ as effectively known and plug in $(\widehat{\rho},\widehat{\gamma}^2)$. Since
\begin{equation*}
    \theta_K = \theta_K^\ast - \phi_K,
    \qquad
    \Delta_K \sim \mathcal{N}(\rho,\gamma^2),
\end{equation*}
a natural debiased point estimate is $\widehat{\theta}_K^\ast - \widehat{\rho}$. Approximating $\widehat{\theta}_K^\ast$ as independent of the latent draw $\Delta_K$ (as is natural when $\Delta_K$ represents cross-domain variability rather than within-domain noise), we obtain the Wald-style interval
\begin{equation}
\mathcal{I}^\ast_{K,1-\alpha}
=
\left[
\widehat{\theta}_K^\ast - \widehat{\rho}
\;\pm\;
z_{1-\alpha/2}\sqrt{\sigma_{\ast,K}^2 + \widehat{\gamma}^2}
\right].
\label{eq:largeK_interval}
\end{equation}
\paragraph{Small $(K-1)$: domain bootstrap over historical domains.}
When $K-1$ is small, uncertainty in $(\widehat{\rho},\widehat{\gamma}^2)$ can materially affect coverage. We therefore propose a \emph{domain bootstrap} that resamples historical domains to propagate this uncertainty. We outline this procedure in Algorithm~\ref{alg:domain_bootstrap}. 
\begin{algorithm}
\caption{Domain-bootstrap confidence interval for $\theta_K$}
\label{alg:domain_bootstrap}
\begin{algorithmic}[1]
\REQUIRE Historical pairs $(\widehat{\theta}_k,\widehat{\theta}_k^\ast)$ and variance components $(\sigma_{0,k}^2,\sigma_{\ast,k}^2,\sigma_{0\ast,k})$ for $k=1,\dots,K-1$; target-domain proxy estimate $\widehat{\theta}_K^\ast$ and $\sigma_{\ast,K}^2$; bootstrap draws $B$; level $\alpha$.
\STATE For each historical domain $k\in\{1,\dots,K-1\}$, compute
\[
d_k := \widehat{\theta}_k^\ast - \widehat{\theta}_k,
\qquad
\widetilde{\sigma}_k^2 := \sigma_{0,k}^2 + \sigma_{\ast,k}^2 - 2\sigma_{0\ast,k}.
\]
\FOR{$b=1$ to $B$}
    \STATE Sample with replacement $K-1$ indices from $\{1,\dots,K-1\}$ to obtain a bootstrap set $\mathcal{K}^{(b)}$.
    \STATE Compute bootstrap moment estimates
    \begin{align*}
    \widehat{\rho}^{(b)}
    &:= \frac{1}{K-1}\sum_{k\in \mathcal{K}^{(b)}} d_k,
    \\
    \widehat{\gamma}^{2\,(b)}
    &:= \frac{1}{K-1}\sum_{k\in \mathcal{K}^{(b)}}(d_k-\overline{d}^{(b)})^2
    \;-\;
    \frac{1}{K-1}\sum_{k\in \mathcal{K}^{(b)}}\widetilde{\sigma}_k^2,
    \end{align*}
    where $\overline{d}^{(b)} := \frac{1}{K-1}\sum_{k\in \mathcal{K}^{(b)}} d_k$, and set
    $\widehat{\gamma}^{2\,(b)} \leftarrow \max\{0,\widehat{\gamma}^{2\,(b)}\}$.
    \STATE Draw a bootstrap replicate of the target primary metric via
    \[
    \theta_K^{(b)} \sim \mathcal{N}\!\left(
    \widehat{\theta}_K^\ast - \widehat{\rho}^{(b)},
    \;\sigma_{\ast,K}^2 + \widehat{\gamma}^{2\,(b)}
    \right).
    \]
\ENDFOR
\STATE Return the empirical $(\alpha/2)$ and $(1-\alpha/2)$ quantiles of $\{\theta_K^{(1)},\dots,\theta_K^{(B)}\}$ as the $(1-\alpha)$ confidence interval for $\theta_K$.
\end{algorithmic}
\end{algorithm}

\section{Simulations: Estimating a Mean under Covariate Shift and Concept Drift}
\label{sec:simulations}
We use a multi-domain simulation to study proxy-based inference when (i) domains differ in covariate distributions (\emph{covariate shift}) and (ii) the proxy--primary relationship varies across domains (\emph{concept drift}). The target estimand is the prevalence of a binary primary outcome in a held-out \emph{target} domain. We evaluate coverage for three proxy-based estimators—an unadjusted proxy mean, an unweighted prediction-powered (PPI) estimator, and a covariate-shift-weighted PPI estimator—both with and without our bias interval adjustment of section~\ref{sec:randombias}. Although the induced proxy bias is not designed to be Gaussian, we find that the Gaussian random-bias model yields well-calibrated coverage across a range of drift magnitudes, while only slightly inflating the confidence intervals when the proxy has correct coverage. 

\subsection{Simulation setup}
\label{sec:sim_setup}

We simulate $K$ domains indexed by $k\in\{1,\dots,K\}$, with domains $\{1,\dots,K-1\}$ serving as labeled \emph{training} domains and domain $K$ serving as the unlabeled \emph{target} domain. In each domain $k$, we generate $n$ i.i.d.\ observations
\[
\{(X_i,Y_i,Y_i^\ast): i\in N_k\},
\qquad
N_k := \{i: S_i=k\},
\]
where $S_i$ is the domain indicator. Covariates satisfy $X_i\in\mathbb{R}^P$ with $P=4$ and are drawn according to
\[
X_i \mid (S_i=k) \sim \mathcal{N}(\mu_k,I_P),
\]
with domain-specific means $\mu_k\in\mathbb{R}^P$. We sample $\mu_1,\dots,\mu_{K-1}$ i.i.d.\ uniformly from the $\ell_2$ unit ball $\{\mu:\|\mu\|_2\le 1\}$ and set $\mu_K=\mathbf{0}$ for the target domain.
\paragraph{Primary outcome with concept drift.}
The primary outcome is binary, $Y_i\in\{0,1\}$. Define the thresholded feature count
\[
T(x;\delta) := \sum_{p=1}^{P}\mathbb{I}\{x_p \ge \delta\} - \frac{P}{2}.
\]
Then for any $x\in\mathbb{R}^P$ and domain $k$,
\[
\mathbb{P}(Y=1 \mid X=x,S=k)
=
\left(
1+\exp\left\{
-\lambda_1\,T(x;\delta_k)+\phi_1
\right\}
\right)^{-1}.
\]
Concept drift is induced through domain-specific thresholds:
\[
\delta_k \sim \mathcal{N}\!\left(0,\frac{\kappa^2}{2}\right)
\quad \text{for } k=1,\dots,K-1,
\qquad
\delta_K:=0.
\]
Thus, $\kappa=0$ corresponds to pure covariate shift (domains differ only through $\mu_k$), while $\kappa>0$ introduces conditional shift through $\delta_k$.
\paragraph{Target estimand.}
The target parameter is the prevalence in the target domain,
\[
\theta_K := \mathbb{E}[Y\mid S=K].
\]
\paragraph{Proxy outcome.}
We generate a misspecified proxy score $Y^\ast\in[0,1]$ as a smooth function of $X$:
\[
Y^\ast
=
\frac{1}{\pi}\arctan\!\left(
\lambda_2 T(x;0)+\phi_2
\right)
+\frac{1}{2}.
\]
When $\kappa=0$, a covariate-shift corrected estimator can (in principle) correct the proxy using labeled domains. When $\kappa>0$, a residual bias will persist regardless of the weighting used. 

To isolate the effect of concept drift (rather than weight estimation error), we use oracle density-ratio weights implied by the Gaussian covariate model. The weight transporting a source domain $k$ to a target domain $k'$ is
\[
w_{k\to k'}(x) := \frac{p_{k'}(x)}{p_k(x)}
=
\exp\!\left(
-\frac{1}{2}\|x-\mu_{k'}\|_2^2
+\frac{1}{2}\|x-\mu_k\|_2^2
\right).
\]
We write $w^{(i)}_{k\to k'}:=w_{k\to k'}(X_i)$ for unit $i\in N_k$.

\subsection{Competing proxy estimates}
\label{subsec:competing_estimators}
We compare the following estimators for $\theta_K$. The primary-only estimator is an oracle benchmark available only in simulation. We will compare against proxy estimators which use only the proxy as well as prediction powered inference estimators \citep{angelopoulos2023prediction}, one vanilla version, and one weighted to adjust for covariate shift. 
\paragraph{Primary-only (oracle).}
\begin{equation}
\widehat{\theta}_K
:=
\frac{1}{n}\sum_{i\in N_K} Y_i.
\label{eq:est_primary_only}
\end{equation}
\paragraph{Proxy-only.}
\begin{equation}
\widehat{\theta}_K^\ast
:=
\frac{1}{n}\sum_{i\in N_K} Y_i^\ast.
\label{eq:est_proxy_only}
\end{equation}
\paragraph{Unweighted PPI} Let $D_i:=Y_i-Y_i^\ast$. Define the cross-domain rectifier
\[
\widehat{\Delta}_{K}
:=
\frac{1}{K-1}\sum_{k=1}^{K-1}
\left(
\frac{1}{n}\sum_{i\in N_k} D_i
\right).
\]
Then the PPI estimator is
\begin{equation}
\widehat{\theta}_K^{\ast,\mathrm{PPI}}
:=
\widehat{\theta}_K^\ast + \widehat{\Delta}_{K}.
\label{eq:est_ppi}
\end{equation}
\paragraph{Weighted PPI (oracle covariate-shift correction).}
For each source domain $k\le K-1$, define normalized within-domain transport weights
\[
\widetilde{w}^{(i)}_{k\to K}
:=
\frac{w^{(i)}_{k\to K}}{\sum_{j\in N_k} w^{(j)}_{k\to K}},
\qquad
\sum_{i\in N_k}\widetilde{w}^{(i)}_{k\to K}=1.
\]
Define the weighted rectifier
\[
\widehat{\Delta}^{(W)}_{K}
:=
\frac{1}{K-1}\sum_{k=1}^{K-1}
\left(
\sum_{i\in N_k}\widetilde{w}^{(i)}_{k\to K}\,D_i
\right),
\]
and the weighted-PPI estimator
\begin{equation}
\widehat{\theta}_K^{\ast,\mathrm{PPI\text{-}W}}
:=
\widehat{\theta}_K^\ast + \widehat{\Delta}^{(W)}_{K}.
\label{eq:est_weighted_ppi}
\end{equation}

Each proxy-based estimator targets a potentially biased estimand. When covariate-shift assumptions hold ($\kappa = 0$) and weights are correct, $\widehat{\theta}_K^{\ast,\mathrm{PPI\text{-}W}}$ targets $\theta_K$; while when $\kappa>0$, $\widehat{\theta}_K$ will become biased ($\theta_K \not = \theta^{\ast,\mathrm{PPI\text{-}W}}_K$). For brevity, we leave the derivations of the covariances of these estimators to the appendix in section~\ref{subsec:cov_components}.

\subsection{Simulation results}

\begin{figure}[t]
    \centering
    \begin{subfigure}[t]{0.95\linewidth}
        \centering
        \includegraphics[width=\linewidth]{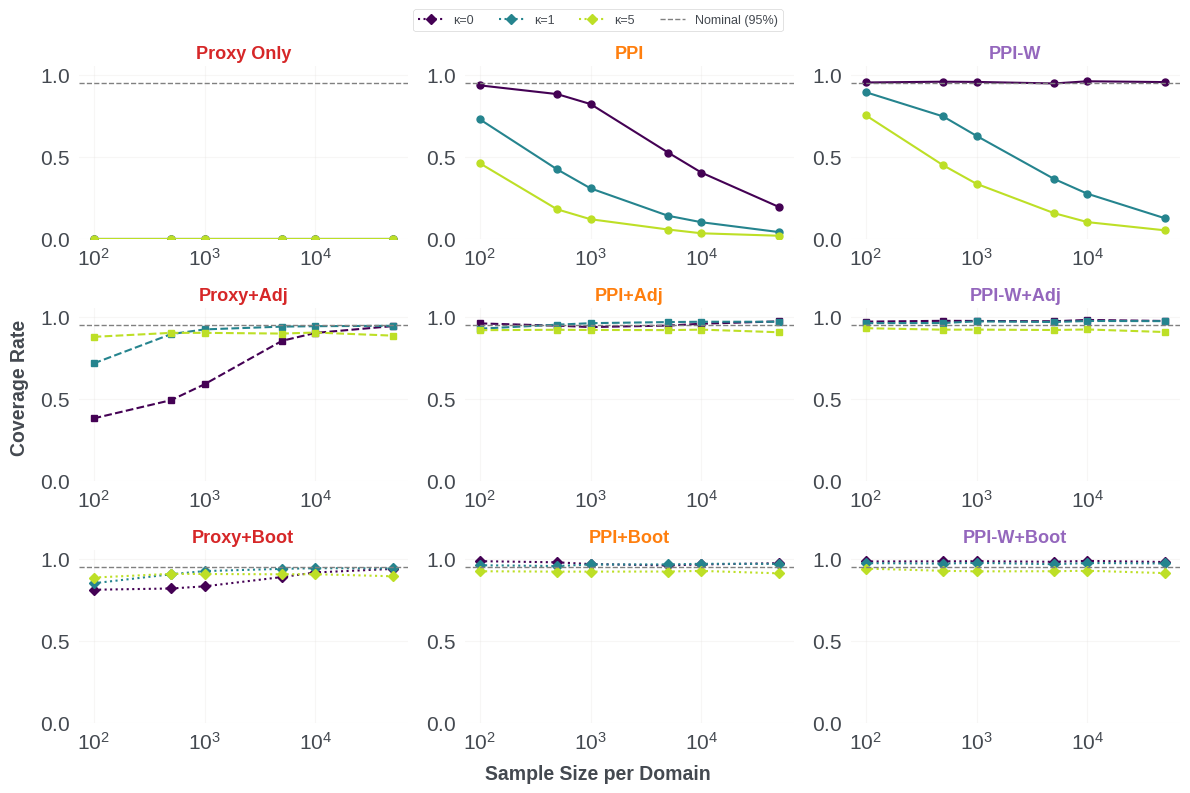}
        \caption{Coverage with $5$ simulated domains.}
        \label{fig:coverage_K5}
    \end{subfigure}
    \vspace{1.0em} 
    \begin{subfigure}[t]{0.95\linewidth}
        \centering
        \includegraphics[width=\linewidth]{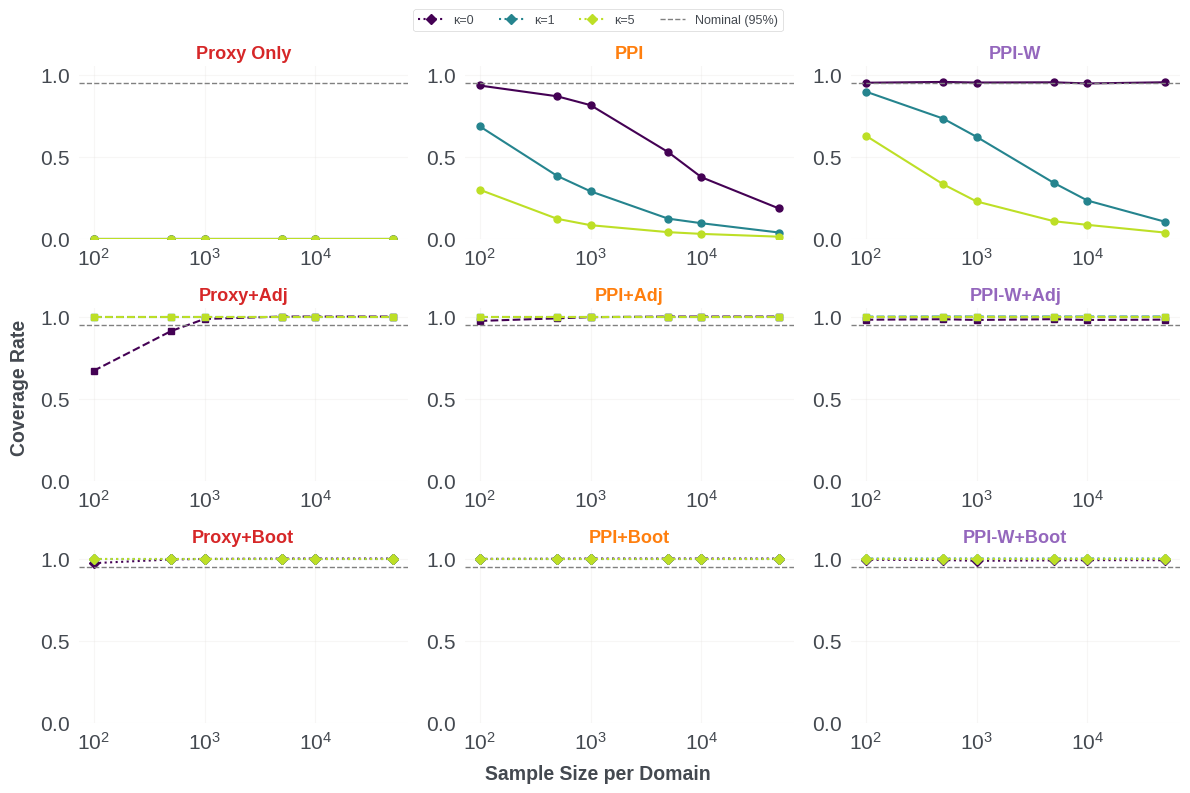}
        \caption{Coverage with $25$ simulated domains.}
        \label{fig:coverage_K25}
    \end{subfigure}
    \caption{Empirical coverage of baseline and adjusted proxy-based intervals across concept drift levels $\kappa \in \{0,1,5\}$. We compare proxy-only, PPI, and weighted PPI, each with no adjustment, plug-in random-bias adjustment, and domain-bootstrap adjustment.}
    \label{fig:coverage_subfigs}
\end{figure}

\paragraph{Coverage.} We compare empirical $(1-\alpha)$ coverage across increasing levels of concept drift, parameterized by $\kappa$ (Figure~\ref{fig:coverage_subfigs}). The proxy-only estimator undercovers across all settings, reflecting persistent residual bias (Figure~\ref{fig:coverage_subfigs}). Vanilla PPI initially improves coverage, but its coverage degrades as the within-domain sample size increases: as sampling uncertainty shrinks, any remaining cross-domain bias which is not properly weighted dominates the error (Figure~\ref{fig:coverage_subfigs}). Incorporating covariate-shift correction via weighted PPI improves calibration, and in the no-drift setting ($\kappa=0$) yields near-nominal coverage throughout (Figure~\ref{fig:coverage_subfigs}). 
Applying our random-bias adjustment further stabilizes coverage as sample size grows (Figure~\ref{fig:coverage_subfigs}). For a small number of historical domains ($K=5$; Figure~\ref{fig:coverage_K5}), the plug-in (non-bootstrap) adjustment can still exhibit dips in severe-bias regimes (especially for proxy-only), because $(\widehat{\rho},\widehat{\gamma}^2)$ are themselves noisy. The domain-bootstrap variant mitigates this effect by propagating uncertainty in the estimated bias distribution (Figure~\ref{fig:coverage_K5}). These finite-$K$ effects are reduced when more historical domains are available ($K=25$; Figure~\ref{fig:coverage_K25}).

\begin{figure}[t]
    \centering
    \includegraphics[width=\linewidth]{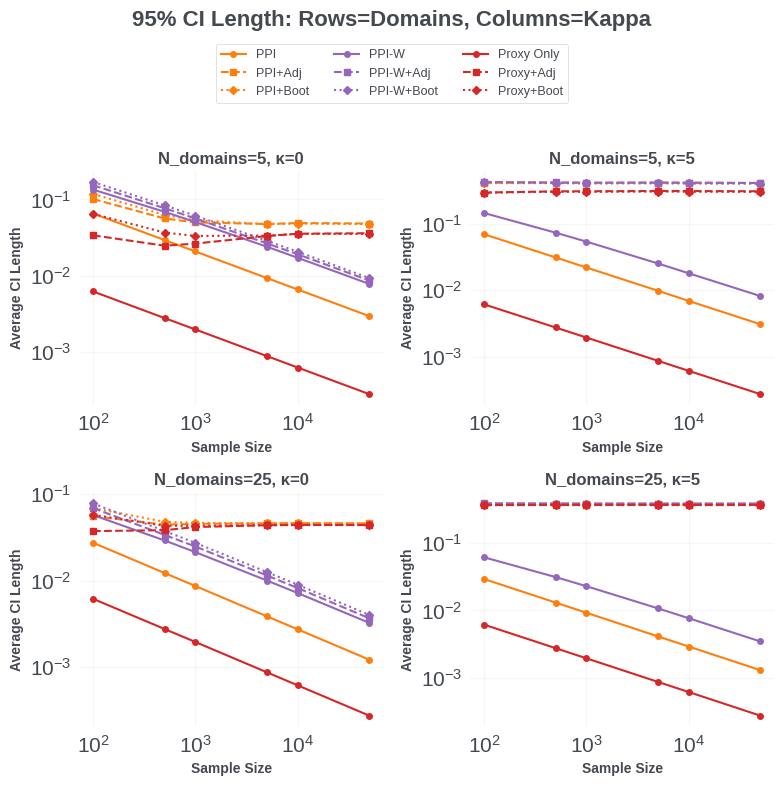}
    \caption{Average confidence-interval length versus within-domain sample size $n$ across proxy estimators and adjustment methods, shown for multiple concept-drift levels $\kappa$.}
    \label{fig:ci_lengths}
\end{figure}

\paragraph{Interval Lengths.} We next compare the lengths of the $(1-\alpha)$ confidence intervals as a function of within-domain sample size $n$ (Figure~\ref{fig:ci_lengths}). When there is covariate shift only ($\kappa=0$) and we use the corrected PPI-W proxy, our bias adjustment intervals yields similar intervals whose lengths shrink with $n$ at the expected parametric rate ($n^{-1/2}$). 

In contrast, for the proxy-only and vanilla PPI proxy estimator, interval lengths eventually plateau as $n$ increases: once sampling error becomes small, residual bias dominates the uncertainty, so additional samples no longer translate into meaningfully shorter calibrated intervals. Under strong concept drift ($\kappa=5$), this bias-dominated regime occurs for all proxy procedures, and interval lengths remain broadly similar across methods, reflecting that the dominant uncertainty is cross-domain bias rather than within-domain sampling variability.

\section{Applications}

To demonstrate the practical utility of our adjustment, we study two real-world applications that differ in data modality, domains, and estimands. First, we consider the \emph{Civil Comments} dataset \cite{borkan2019nuanced}, available through the WILDS benchmark \cite{koh2021wilds}, where the goal is to estimate the prevalence of toxic comments across web domains. Second, we analyze a collection of controlled experiments, where the estimand is the average treatment effect on a long term outcome. Together, these applications illustrate that the proposed adjustment can be applied across distinct settings and target parameters.  Table~\ref{tab:results_application} summarizes performance using two complementary evaluation criteria computed over historical domains: (i) hold-one-out \emph{interval overlap} at level $\alpha=0.05$ as described in section~\ref{subsec:evaluation} and (ii) \emph{normalized average interval width}, the ratio of the confidence interval width of the proxy compared to the (oracle) primary outcomes. 

\begin{table}[t]
\centering
\caption{Adjustment Results for Real Datasets}
\label{tab:results_application}
\small
\setlength{\tabcolsep}{4pt} 
\begin{tabularx}{\linewidth}{@{}lllc c@{}}
\toprule
Dataset & Metric & Method & \shortstack{Overlap \\($\alpha=0.05$)} & \shortstack{Avg.\ Interval\\Width (norm.)} \\
\midrule
CC & Prevalence & Proxy       & 0.64 & 0.5 \\
CC & Prevalence & Proxy + Adj & 0.96 & 1.3 \\
\midrule
LTE & ATE & Proxy       & 0.68 & 1.2 \\
LTE & ATE & Proxy + Adj & 1 & 2.9 \\
\bottomrule
\end{tabularx}
\vspace{0.5em}
\parbox{\linewidth}{\footnotesize\textit{Note:} CC = Civil Comments; LTE = Long Term Experiment; ATE = Average Treatment Effect}
\end{table}

\subsection{Civil Comments}
The Civil Comments dataset consists of user-generated comments on online articles annotated with a binary toxicity label \cite{borkan2019nuanced}. Each comment is associated with the publication in which the corresponding article appeared; we treat publication as the domain variable. Comment counts are imbalanced across publications: across all publications, the largest domain contains 27{,}077 comments while the smallest contains 13, with an average of roughly 3{,}000 comments per publication. For stable domain-level inference, we restrict attention to the 25 publications with at least 10 comments.

We construct a proxy outcome by training a text classifier to predict toxicity from comment text. Concretely, we use a BERT-based encoder with a maximum sequence length of 128 tokens. We attach a single linear classification head that maps the transformer hidden representation (e.g., the pooled \texttt{[CLS]} embedding) to a scalar logit. The model is trained using binary cross-entropy with logits and the AdamW optimizer. We then apply the trained model to the held-out test set to obtain proxy scores $Y_i^\ast \in [0,1]$ for each comment.

Our target estimand is the prevalence of toxic comments within each publication. Treating publications as domains, we apply our interval adjustment procedure by using historical publications (i.e., other domains with observed toxicity labels) to estimate the residual proxy bias distribution and then inflate proxy-based uncertainty accordingly.

\begin{figure}[t]
    \centering
    \begin{subfigure}[t]{0.95\linewidth}
        \centering
        \includegraphics[width=\linewidth]{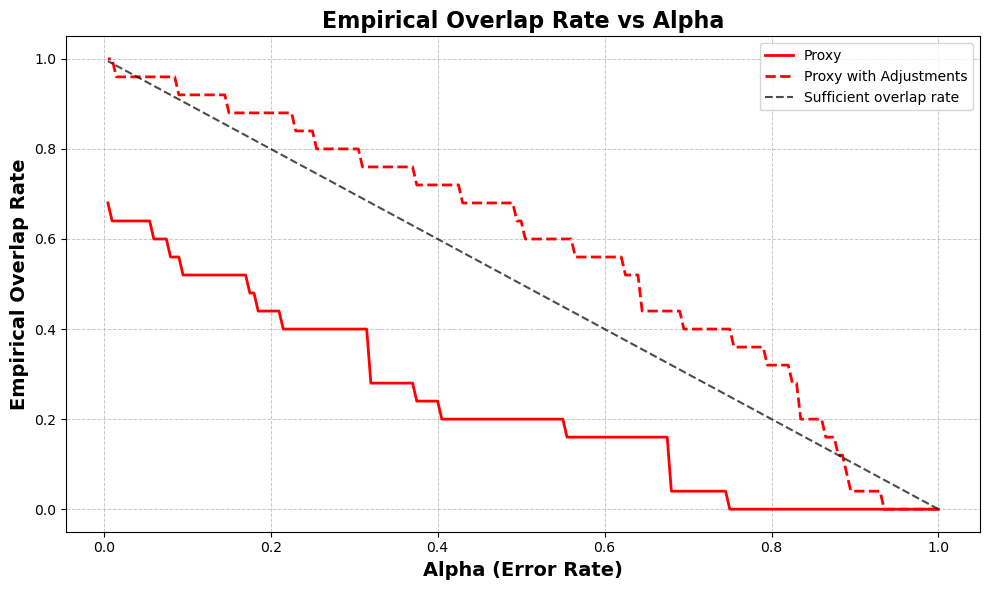}
        \caption{Interval overlap rate versus nominal confidence level.}
        \label{fig:overlap_rate_toxic_comments}
    \end{subfigure}
    \vspace{1.0em} 
    \begin{subfigure}[t]{0.95\linewidth}
        \centering
        \includegraphics[width=\linewidth]{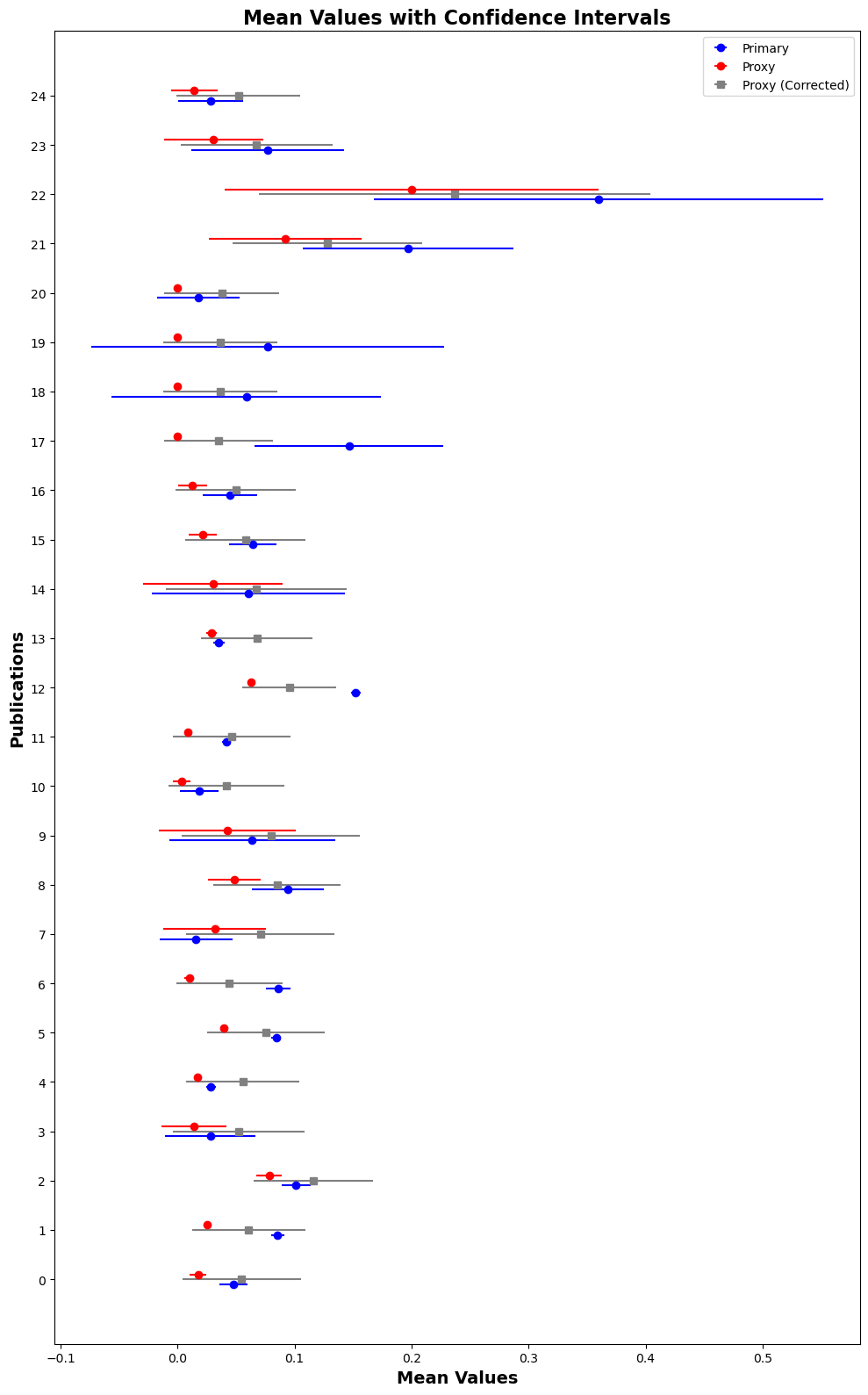}
        \caption{Publication-level point estimates and confidence intervals.}
        \label{fig:confidence_intervals_toxic_comments}
    \end{subfigure}
    \caption{Civil Comments: publication-level prevalence estimation using proxy-only and proxy-with-adjustment procedures.}
    \label{fig:toxic_comments_results}
\end{figure}

Figure~\ref{fig:toxic_comments_results} summarizes the results. Across confidence levels, the overlap rate of proxy intervals with adjustment is closer to nominal than that of proxy-only intervals (Figure~\ref{fig:toxic_comments_results}\subref{fig:overlap_rate_toxic_comments}). The adjusted proxy intervals also align more closely with primary-metric intervals across publications by inflating uncertainty in a domain-adaptive way (Figure~\ref{fig:toxic_comments_results}\subref{fig:confidence_intervals_toxic_comments}). At $\alpha=0.05$, overlap increases from $0.64$ (proxy-only) to $0.96$ (proxy + adjustment), while the normalized average interval length increases from $0.5$ to $1.3$, reflecting the cost of improved calibration in this setting.

\subsection{Long Term Experiment}

We next apply our adjustment procedure to a collection of controlled experiments. For each experiment, the primary estimand is the average treatment effect (ATE) which takes time to manifest. In order to run experiments more readily, analysts often rely on a proxy outcome: a model-predicted version of the long-term outcome. While this proxy enables faster iteration, proxy-based inference can be miscalibrated due to prediction error and residual distribution shift (i.e., a changing relationship between predicted and realized long-term outcome across experiments).

\begin{figure}[t]
    \centering
    \includegraphics[width=\linewidth]{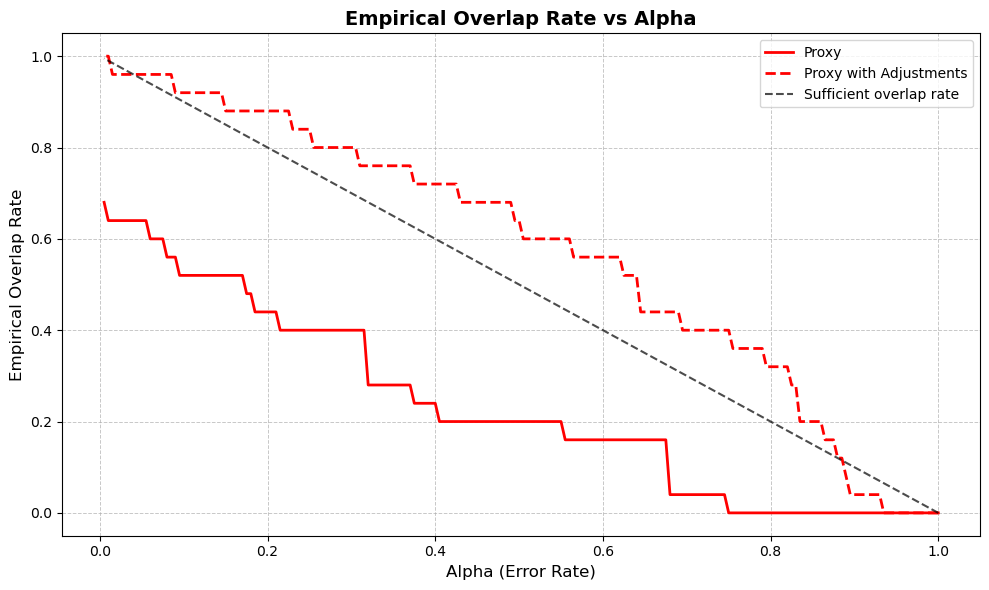}
    \caption{Long term experiments: empirical interval overlap rate versus nominal confidence level for proxy-only and proxy-with-adjustment procedures.}
    \label{fig:overlap_long_term_outcome}
\end{figure}

We use our method to construct confidence intervals for the experiment-level ATE using proxy-based estimates, while explicitly accounting for residual proxy bias learned from historical experiments. Empirically, the adjustment inflates uncertainty when warranted and improves agreement with primary-metric intervals. Figure~\ref{fig:overlap_long_term_outcome} reports overlap rates across confidence levels: proxy-only intervals tend to be overconfident, whereas proxy + adjustment yields overlap much closer to nominal. At $\alpha=0.05$, overlap increases from $0.68$ (proxy-only) to $1$ (proxy + adjustment), while normalized average interval width increases from $1.2$ to $2.9$ (Table~\ref{tab:results_application}). In this application, the gains are most pronounced in experiments where residual proxy bias is large relative to the proxy estimator's sampling variability, so that naive proxy intervals understate uncertainty.

Overall, this case study illustrates how the proposed adjustment supports faster decision-making with proxy outcomes while maintaining rigorous uncertainty quantification. By reducing false positives from overconfident proxy inference, the method can allow for safer experiment cycles in settings where incorrect launches can be harmful at scale.

\section{Discussion} 

This paper studies inference when the primary outcome is unavailable in a target domain and analyses must rely on a proxy metric. We propose an estimate-level adjustment that inflates proxy-based uncertainty to account for residual proxy bias learned from historical domains, using only aggregate proxy/primary estimates and covariance summaries. Because the adjustment is modular, it can be layered on top of existing proxy estimators and correction methods (e.g., prediction-powered estimators): when identification assumptions are approximately valid the adjustment is minimal, and when assumptions fail it becomes appropriately conservative, improving empirical coverage rate. Ultimately we can think of this method as a post-processing step on existing proxies which safely adjusts the intervals. 

A primary limitation of this method is that it improves uncertainty quantification but does not directly improve the proxy itself, so in strongly bias-dominated regimes intervals may remain wide if the proxy is poor. Its usefulness also depends on the extent to which historical domains are informative about the target; a drastic change in the target domain, which changes the relationship of the primary to the proxy unlike what was seen in historical domains, cannot be corrected. In addition, when the number of historical domains is small, the estimated bias distribution is noisy; bootstrap variants help but do not eliminate the scarcity issue. Finally, the Gaussian random-bias model is a working approximation, and heavier-tailed or multi-modal bias patterns may require richer, nonparametric alternatives.

These limitations motivate several open questions: how best to share bias information across heterogeneous domain types and under gradual drift; how to select or combine multiple candidate proxies to optimize a calibrated objective \cite{tripuraneni2024choosing}; and which types of model properties; such as multicalibration and related robustness notions, lead to accurate proxy metrics \cite{kim2019multiaccuracy,kim2022universal}. We leave an investigation of these aspects of proxy adjustment as future work.

\bibliographystyle{ACM-Reference-Format}
\bibliography{references}

\appendix

\section{Simulation parameters and ground-truth prevalence}
\label{app:sim_params_truth}
In all simulations we set $P=4$. For $s \in \{1,\dots,K-1\}$, we draw $\mu_s$ uniformly from the Euclidean ball $\{\mu \in \mathbb{R}^4 : \|\mu\|_2 \le 1\}$, and we fix the target mean to
\[
\mu_K = \left(\tfrac{1}{2}, -\tfrac{1}{2}, \tfrac{1}{2}, -\tfrac{1}{2}\right).
\]
We vary $n_s \in \{100,500,1000,5000, 10000, 50000\}$, number of domains: $K \in \{5,10,25\}$, and concept drift: $\kappa \in \{0,0.01,0.1,1,10\}$. We use $\lambda_1=\tfrac{1}{2}$, $\phi_1=2$ in the outcome model and $\lambda_2=\tfrac{1}{2}$, $\phi_2=2$ in the proxy model. For concept drift, for each source domain $s<K$ we draw
\[
\delta_s \sim \mathcal{N}\!\left(0,\kappa^2/2\right),
\]
and we keep the target threshold $\delta_K$ fixed (so that the target prevalence is comparable across Monte Carlo replicates). For each $(n_s,K,\kappa)$ configuration, we run $1000$ independent replicates and compute empirical coverage and average interval length.
To evaluate coverage, we approximate the true target prevalence $\theta=\mathbb{E}[Y\mid S=K]$ by Monte Carlo with $M=10^6$ samples:
\[
X^{(m)} \stackrel{iid}{\sim} \mathcal{N}(\mu_K,I_4), \qquad
\widehat{\theta}_{MC}=\frac{1}{M}\sum_{m=1}^M \mathbb{P}(Y=1 \mid X=X^{(m)}, S=K),
\]
where the conditional probability uses the target-domain parameters (including the fixed $\delta_K$). We treat $\widehat{\theta}_{MC}$ as ground truth when computing empirical coverage.

\subsection{Estimating covariance components of proxy estimators.}
\label{subsec:cov_components}
We estimate the within-domain covariance matrices required by the random-bias procedure using standard sample covariance formulas.
\paragraph{Within-domain covariance of primary and proxy means (labeled domains).}
For a labeled domain $k\le K-1$, define the primary and proxy means, which are components of the estimators
\[
\bar Y_k := \frac{1}{n}\sum_{i\in N_k} Y_i,
\qquad
\bar Y_k^\ast := \frac{1}{n}\sum_{i\in N_k} Y_i^\ast.
\]
Denote the sample variances
\[
S_{k,YY}:=\frac{1}{n-1}\sum_{i\in N_k}(Y_i-\bar Y_k)^2,
\quad
S_{k,Y^\ast Y^\ast}:=\frac{1}{n-1}\sum_{i\in N_k}(Y_i^\ast-\bar Y_k^\ast)^2,
\]
\[
S_{k,YY^\ast}:=\frac{1}{n-1}\sum_{i\in N_k}(Y_i-\bar Y_k)(Y_i^\ast-\bar Y_k^\ast).
\]
Then
\[
\widehat{\mathrm{Cov}}
\!\left(
\begin{bmatrix}
\widehat{\theta}_k\\
\widehat{\theta}_k^\ast
\end{bmatrix}
\right) = 
\widehat{\mathrm{Cov}}
\!\left(
\begin{bmatrix}
\bar Y_k\\
\bar Y_k^\ast
\end{bmatrix}
\right)
=
\frac{1}{n}
\begin{bmatrix}
S_{k,YY} & S_{k,YY^\ast}\\
S_{k,YY^\ast} & S_{k,Y^\ast Y^\ast}
\end{bmatrix}.
\]
\paragraph{Variance of the PPI rectifier (unweighted).}
Let $D_i:=Y_i-Y_i^\ast$ and $\bar D_k:=\frac{1}{n}\sum_{i\in N_k}D_i$. Under cross-domain independence,
\[
\text{Var}(\widehat{\Delta}_K)
=
\frac{1}{(K-1)^2}\sum_{k=1}^{K-1}\text{Var}(\bar D_k),
\qquad
\widehat{\text{Var}}(\bar D_k)=\frac{S_{k,DD}}{n},
\]
where
\[
S_{k,DD}:=\frac{1}{n-1}\sum_{i\in N_k}(D_i-\bar D_k)^2.
\]
Hence,
\[
\widehat{\text{Var}}(\widehat{\Delta}_K)
=
\frac{1}{(K-1)^2}\sum_{k=1}^{K-1}\frac{S_{k,DD}}{n}.
\]

Furthermore, note that $\text{Cov}(\bar Y_k, \widehat{\Delta}_k) = \text{Cov}(\bar Y_k^\ast, \widehat{\Delta}_k) = 0$ since the observational units in the rectifier $\widehat{\Delta}_k$ do not overlap with $\bar Y_k$. 

\paragraph{Variance of the weighted rectifier.}
For each source domain $k$, define the weighted residual mean
\[
\widehat{\Delta}^{(W)}_k
:=
\sum_{i\in N_k}\widetilde{w}^{(i)}_{k\to K}\,D_i,
\qquad
\widehat{\Delta}^{(W)}=\frac{1}{K-1}\sum_{k=1}^{K-1}\widehat{\Delta}^{(W)}_k.
\]
Again using cross-domain independence,
\[
\text{Var}(\widehat{\Delta}^{(W)})
=
\frac{1}{(K-1)^2}\sum_{k=1}^{K-1}\text{Var}(\widehat{\Delta}^{(W)}_k),
\]
and we estimate the within-domain variance by the plug-in weighted-mean formula
\[
\widehat{\text{Var}}(\widehat{\Delta}^{(W)}_k)
=
\sum_{i\in N_k}(\widetilde{w}^{(i)}_{k\to K})^2\,
\bigl(D_i-\widehat{\Delta}^{(W)}_k\bigr)^2.
\]
\paragraph{Variance of proxy-based estimators in the target domain.}
Because rectifiers are computed from source domains, they are independent of the target-domain proxy mean $\bar Y_K^\ast=\frac{1}{n}\sum_{i\in N_K}Y_i^\ast$ under cross-domain independence. Therefore,
\begin{align*}
    \text{Var}(\widehat{\theta}_K^{\ast,\mathrm{PPI}}) &= \text{Var}(\bar Y_K^\ast)+\text{Var}(\widehat{\Delta}),
    \text{Var}(\widehat{\theta}_K^{\ast,\mathrm{PPI\text{-}W}}) &= \text{Var}(\bar Y_K^\ast)+\text{Var}(\widehat{\Delta}^{(W)}),
\end{align*}

with $\text{Var}(\bar Y_K^\ast)$ estimated by $S_{K,Y^\ast Y^\ast}/n$ (using the sample variance of $Y^\ast$ in the target domain).

\section{Methodology Details and Extensions}


\subsection{Model Extensions: Contextual Modeling}
\label{subsec:contextual}
In some applications, domains may naturally be grouped into types. Furthermore, these types may be continuous if a model is monitored across time, we may look at the residual biased based on nearby time periods. 
\paragraph{Setup.}
Let each historical domain $k\in\{1,\dots,K-1\}$ have observed context $C_k \in \mathcal{C}$ (this may be a type of experiment, or a time indicator). We propose now that the domain bias has a 
\[
\phi_k = \theta_k - \theta_k^*  \;\sim\; \mathcal{N}\!\left(\rho(C_k),\; \gamma^2(C_k) + \widetilde{\sigma}_k^2\right).
\]
\paragraph{Similarity-weighted sharing.}
To estimate $(\rho(c),\gamma^2(c))$ for a target context value $c\in\mathcal{C}$, define a nonnegative similarity function $S(c,c';\beta)$ and normalize it into a set of domain weights $v_k$:
\[
v_k(c;\beta)
:= \frac{S(c,C_k;\beta)}{\sum_{k'=1}^{K-1} S(c,C_{k'};\beta)},
\qquad k=1,\dots,K-1.
\]


\paragraph{Weighted method-of-moments.}
Using the weights, we define context-specific moment estimators:
\begin{align}
\widehat{\rho}(c)
&:= \sum_{k=1}^{K-1} v_k(c;\beta)\, d_k,
\label{eq:rho_context}
\\
\widehat{\gamma}^2(c)
&:= \sum_{k=1}^{K-1} v_k(c;\beta)\,\bigl(d_k-\widehat{\rho}(c)\bigr)^2
\;-\;
\sum_{k=1}^{K-1} v_k(c;\beta)\,\widetilde{\sigma}_k^2, \\
\widehat{\gamma}^2(c) &\leftarrow \max\{0,\widehat{\gamma}^2(c)\}.
\label{eq:gamma_context}
\end{align}
These estimators reduce to the unweighted moments when $w_k(c;\beta)=1/(K-1)$.
\paragraph{Context-specific target-domain interval.}
If the target domain has context $C_K=c$, we plug in $(\widehat{\rho}(c),\widehat{\gamma}^2(c))$ in the large-$(K-1)$ interval:
\[
\mathcal{I}^\ast_{K,1-\alpha}(c)
=
\left[
\widehat{\theta}_K^\ast - \widehat{\rho}(c)
\;\pm\;
z_{1-\alpha/2}\sqrt{\sigma_{\ast,K}^2 + \widehat{\gamma}^2(c)}
\right].
\]
If each domain has a timestamp $t_k$ and drift is expected to be smooth, incorporate a time-decay kernel $K_h(\cdot)$ (e.g., exponential or Gaussian) by replacing
\[
S(c,C_k;\beta) \quad \text{with} \quad S(c,C_k;\beta)\,K_h(t_K-t_k),
\]
and re-normalizing to obtain weights. This emphasizes more recent domains when estimating $(\rho(c),\gamma^2(c))$.

A pragmatic tuning rule is to choose $\beta$ by grid search to maximize a weighted Gaussian marginal likelihood for $\{d_k\}$ under the working model:
\[
\ell(\beta;c)
:=
-\frac{1}{2}\sum_{k=1}^{K-1} w_k(c;\beta)\left[
\log\!\bigl(\widehat{\gamma}^2(c;\beta)+\widetilde{\sigma}_k^2\bigr)
+
\frac{\bigl(d_k-\widehat{\rho}(c;\beta)\bigr)^2}{\widehat{\gamma}^2(c;\beta)+\widetilde{\sigma}_k^2}
\right],
\]
where $(\widehat{\rho}(c;\beta),\widehat{\gamma}^2(c;\beta))$ are computed from~\eqref{eq:rho_context}--\eqref{eq:gamma_context} at that $\beta$.

\subsection{Segmentation, or multivariate metrics}
In many instances, we would like to condition on types of domains, as well as segment results by discrete covariates. Let $x\in\mathcal{X}$ denote a discrete segment label (e.g., country, user type). Let $\theta_{k,x}$ and $\theta_{k,x}^\ast$ denote the primary and proxy parameters within segment $x$ in domain $k$, and let $\boldsymbol{\theta}_k := (\theta_{k,x})_{x\in\mathcal{X}}$ and $\boldsymbol{\theta}_k^\ast := (\theta_{k,x}^\ast)_{x\in\mathcal{X}}$.
A multivariate extension models
\begin{equation}
\boldsymbol{\theta}_k^\ast - \boldsymbol{\theta}_k =: \boldsymbol{\epsilon}_k \sim \mathcal{N}(\boldsymbol{\rho},\mathbf{\Gamma}),
\qquad k=1,\dots,K,
\label{eq:seg_multivar_bias}
\end{equation}
and assumes an asymptotic approximation
\[
\widehat{\boldsymbol{\theta}}_k^\ast - \boldsymbol{\theta}_k^\ast =: \boldsymbol{\delta}_k \sim \mathcal{N}(\mathbf{0},\mathbf{\Sigma}_k),
\]
where $\mathbf{\Sigma}_k$ is the covariance matrix of the segment-wise proxy estimators. If segments are disjoint, $\mathbf{\Sigma}_k$ is diagonal. Then
\[
\widehat{\boldsymbol{\theta}}_k^\ast - \boldsymbol{\theta}_k
=
\boldsymbol{\delta}_k + \boldsymbol{\epsilon}_k
\sim
\mathcal{N}(\boldsymbol{\rho},\mathbf{\Gamma}+\mathbf{\Sigma}_k).
\]
Since $\Gamma$ is typically high-dimensional, estimation of this covariance can be done using factor analysis approaches. 

\end{document}